\documentclass[twocolumn,showpacs,preprintnumbers,amsmath,amssymb,prb,superscriptaddress]{revtex4}
\usepackage{graphicx}
\usepackage{bm}
\newcommand{\bs}   {\boldsymbol}  
\newcommand{\mb}   {\mathbf}      
\newcommand{\mr}   {\mathrm}

\newcommand{\imag} {\mathrm{i}}   
\newcommand{\dd}   {\mathrm{d}}   
\newcommand{\e}    {\mathrm{e}}   
\newcommand{\bra}  {\langle}      
\newcommand{\ket}  {\rangle}      
\newcommand{\up}   {\uparrow}     
\newcommand{\dn}   {\downarrow}   
\newcommand{\s}    {\sigma}       
\newcommand{\w}    {\omega}       

\newcommand{\eps}  {\epsilon} 
\begin{document}
\preprint{}
\title{A BCS-BEC crossover in the extended Falicov-Kimball model:\\ Variational cluster approach}

\author{K. Seki}
\email{seki-kazuhiro@graduate.chiba-u.jp}
\affiliation{Department of Physics, Chiba University, Chiba 263-8522, Japan}
\author{R. Eder}
\affiliation{Karlsruhe Institute of Technology, Institut f{\"u}r Festk{\"o}rperphysik, D-76021 Karlsruhe, Germany}
\author{Y. Ohta}
\affiliation{Department of Physics, Chiba University, Chiba 263-8522, Japan}

\begin{abstract}
We study the spontaneous symmetry breaking of the excitonic insulator state induced by the Coulomb interaction $U$
in the two-dimensional extended Falicov-Kimball model.
Using the variational cluster approximation (VCA) and Hartree-Fock approximation (HFA), we evaluate 
the order parameter,
single-particle excitation gap, 
momentum distribution functions,
coherence length of excitons, and 
single-particle and anomalous excitation spectra, 
as a function of $U$ at zero temperature.
We find that in the weak-to-intermediate coupling regime, 
the Fermi surface plays an essential role and 
calculated results can be understood 
in close correspondence with the BCS theory, 
whereas in the strong-coupling regime, 
the Fermi surface plays no role and 
results are consistent with the picture of BEC.
Moreover, we find that HFA works well both in the weak- and strong-coupling regime,
and that the difference between the results of VCA and HFA mostly appears in the intermediate-coupling regime.
The reason for this is discussed from a viewpoint of the self-energy. 
We thereby clarify the excitonic insulator state that typifies 
either a BCS condensate of electron-hole pairs (weak-coupling regime) or 
a Bose-Einstein condensate of preformed excitons (strong-coupling regime).
\end{abstract}

\date{\today}
\maketitle


\section{Introduction}


The realization of excitonic insulators (EI) on the proximity of the semimetal-semiconductor 
transition was suggested about half a century ago.~\cite{Mott,Halperin,Jerome}  
Because of the weak screening of the Coulomb attraction between the electrons and holes 
due to the small number of carriers, 
the electrons and holes may spontaneously form bound states (excitons), giving rise to the EI state.  
As a candidate for EI, quasi-one-dimensional 1$T$-TiSe$_2$ has been studied both theoretically and experimentally.~\cite{Cercellier,Monney} 
TaNi$_2$Se$_5$ has also been studied by angle-resolved photoemission spectroscopy (ARPES) measurements. 
It was reported that the valence-band top is extremely flat, 
and the material can be a new candidate for an EI of bound pairs between Ni 3$d$-Se 4$p$ holes and Ta $5d$ electrons.~\cite{Wakisaka,Kaneko}  

From the theoretical point of view, the Falicov-Kimball model~\cite{FalicovKimball} extended by including a finite valence bandwidth, 
i.e., extended Falicov-Kimball model (EFKM), has been extensively studied in context of the EI or electric ferroelectricity.~\cite{Batista1,Batista2}
The EFKM contains 
the large bandwidth $c$-electrons (with a hopping integral $t_c$ and on-site energy $\eps_c$), 
small bandwidth $f$-electrons (with a hopping integral $t_f$ and on-site energy $\eps_f$), 
and a Coulomb interaction ($U$) between $c$- and $f$-electrons. 
The ground state phase diagram of the EFKM in the weak-to-intermediate-coupling regime 
was obtained by the constrained path Monte Carlo (CPMC).~\cite{Batista1}
In the strong-coupling regime, the EFKM can be mapped onto the spin-$1/2$ Ising-like XXZ model with a uniform magnetic field. 
In that case, the spontaneous EI ordering corresponds to the spontaneous magnetization in the XY-plane and its phase diagram was also determined.~\cite{Batista2}
The phase diagram of the EFKM is composed of three phases: 
the charge-density-wave (CDW) with staggered orbital order (SOO) phase, excitonic insulator (EI) phase, and band insulator (BI) phase.
The CDW phase is characterized by the periodic modulation of the total density of $c$- and $f$-electrons, while 
the SOO phase is characterized by the periodic modulation in the difference between the $c$- and $f$-electron densities.
The instability toward the CDW and SOO phases was studied in detail by Zenker \textit{et al.}~\cite{Zenker1} 
The EI phase is characterized by the spontaneous $c$-$f$ hybridization. 
The BI phase is characterized by the completely filled $c$- or $f$-band.
Interestingly, the ground state phase diagram in the weak-to-intermediate-coupling regime obtained by a Hartree-Fock approximation (HFA)~\cite{Farkasovsky} 
agrees quite well with that obtained by CPMC. 
On the other hand, excitation properties of EFKM are still of great interest.
Finite-temperature phase diagram and electron-hole bound state formation in EFKM were studied by HFA.~\cite{Ihle}  
Projector-based remormalization method (PRM) calculation on the one-dimensional EFKM~\cite{Phan1} reported 
that incoherent parts of the single-particle excitation spectra, which are related to the dissociation of the excitons, appears especially in the BEC regime.
Detailed studies on the dynamical excitonic susceptibility at finite-temperature calculated by use of the PRM~\cite{Phan2} and slave boson (SB) technique~\cite{Zenker1} 
confirmed that tightly bound excitons exist even above the critical temperature for  exciton condensation.
The results strongly support the so-called excitonic halo suggested by Bronold and Fehske,~\cite{Bronold}
where tightly bound excitons exist without condensation,
and the scenario of the Bose-Einstein condensation of preformed excitons in the semiconductor side.   
Thus the effects of electron correlations on the static and dynamic properties of this model are worth studying.  

In this paper, we study the EI state of the EFKM defined on the two-dimensional square 
lattice as a function of the Coulomb interaction strength $U$.
We employ the variational cluster approximation (VCA)~\cite{Potthoff1} based 
on the self-energy functional theory (SFT)~\cite{Potthoff2} at zero temperature.
The cluster perturbation theory (CPT)~\cite{Senechal2} is used to calculate the single-particle Green's functions.
We also employ HFA  to clarify the effects of electron correlation that can be taken into account in VCA. 
As far as we know, VCA has not been applied to the study of the EI state of EFKM.
The advantage of VCA compared to HFA is that VCA can fully take into account 
static and dynamic effects of electron correlations within the range of a finite size cluster. 
So far, VCA and CPT applied to a variety of strongly correlated electron systems,
such as the half-filled Hubbard model with competing magnetic orders in two-dimensional~\cite{Senechal3} and three-dimensional systems,~\cite{Yoshikawa} 
periodic Anderson model with the competition between magnetic ordering induced by the Ruderman-Kittel-Kasuya-Yoshida interaction and nonmagnetic Kondo screening,~\cite{Horiuchi} 
multi-orbital system with spin-orbit coupling, XY-plane magnetic ordering,~\cite{Watanabe} etc, 
and it turns out that the method is useful to discuss correlation effects on the symmetry breaking or single-particle excitation spectra,
especially in the insulating state. 
Thus it is worth studying the EI state of the EFKM by applying VCA and CPT to investigate the effects of short range correlations 
on the symmetry breaking or single-particle excitations.

We will first discuss the $U$ dependence of the calculated EI order parameter and single-particle gap.
In the weak-coupling regime, 
as expected from the BCS theory, the single-particle gap is scaled well by the order parameter,
whereas in the strong-coupling regime, 
the order parameter rapidly decreases with increasing $U$.
Then we will show the calculated momentum distribution functions as functions of $U$. 
In the weak-coupling regime, 
the momentum distribution functions behave like those in the BCS theory, 
whereas in the strong-coupling regime 
the momentum dependence of the momentum distribution functions becomes weak, 
and the behavior is quite different from BCS theory.
The coherence length of the exciton shows a shallow minimum at the crossover regime as a function of $U$.
We further calculate the single-particle spectra, anomalous Green's functions, and density of states  
in order to investigate the electron correlation effects on the quasi-particles.
%
Thus our study will shed light on the BCS-BEC crossover~\cite{Nozieres} in the EI state.

This paper is organized as follows.
In Sec.~\ref{sec_method}, we introduce our model and method of calculation.
In Sec.~\ref{sec_result}, we present our results for 
the EI order parameter, 
single-particle gap, 
single-particle Green's function,
anomalous Green's function,
momentum distribution functions,
and coherence length as functions of the Coulomb interaction strength $U$.
Discussion on the efficiency of the HFA on this model and 
experimental implications are given in Sec.~\ref{sec_discussion}.  
We summarize our work in Sec.~\ref{sec_summary}.


\section{Model and Method}
\label{sec_method}



\subsection{Extended Falicov-Kimball model}

The Hamiltonian of the EFKM reads
\begin{eqnarray}\label{eq.ham}
{\cal H} &=& 
   - t_c \sum_{\langle ij \rangle} (c_i^{\dag} c_j + \mathrm{H.c.}) + (\eps_c - \mu) \sum_{i} n_{ic} \nonumber \\
&& - t_f \sum_{\langle ij \rangle} (f_i^{\dag} f_j + \mathrm{H.c.}) + (\eps_f - \mu) \sum_{i} n_{if} \nonumber \\
&&+ U \sum_{i} n_{i c} n_{i f}
\end{eqnarray}
where $c_i$ ($c_i^\dag$) denotes the annihilation (creation) operator of an electron on the $c$-orbital at site $i$ and 
$n_{ic} = c_{i}^\dag c_{i}$.
$t_c$ is the hopping integral between neighboring sites of the two-dimensional square lattice and 
$\eps_c$ is the on-site energy level of the $c$-orbitals.  
These are the same for the $f$-orbitals.
$U$ is the inter-orbital Coulomb repulsion between electrons.
The chemical potential $\mu$ is determined so as to maintain the average particle density $n$ at half filling $n = 1$.  
Throughout the paper, we set $\hbar = k_{B} = 1$ and lattice constant $a=1$. 
We use $t_c = 1$ as the unit of energy and we focus on the band parameter values $\eps_c = 0, \eps_f = -1,$ and $t_f=-0.3$. 

\begin{figure}[t]
\begin{center}
\includegraphics[width=18pc]{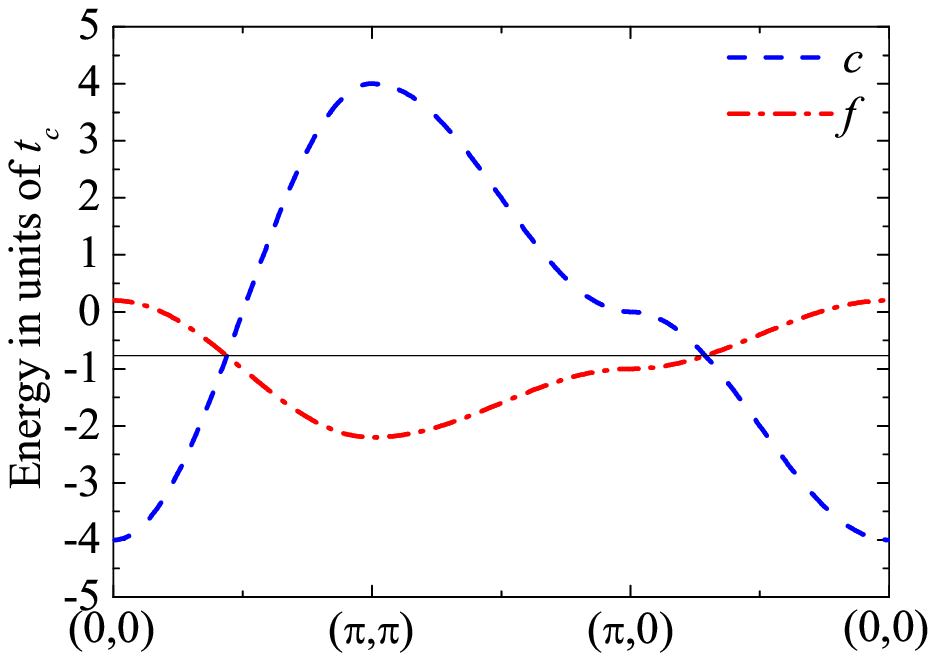}\\
\includegraphics[width=12pc]{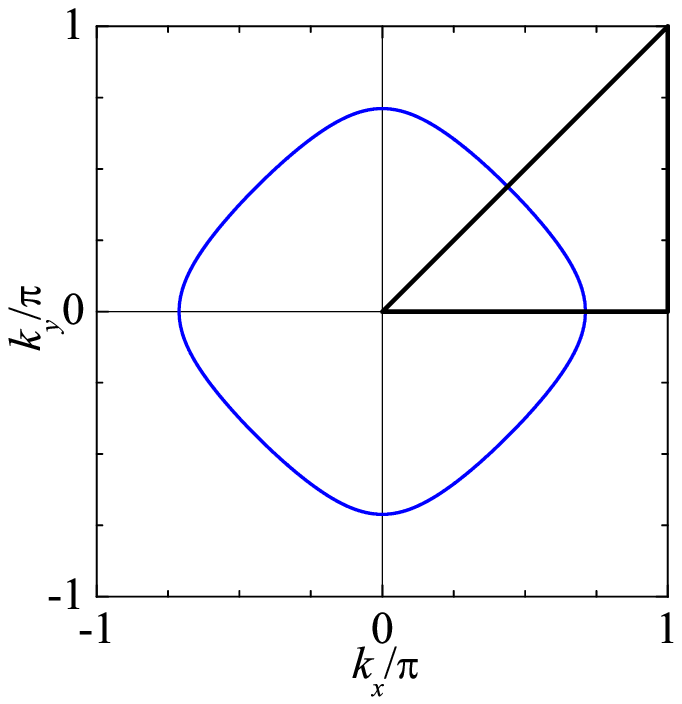}
\caption{
\label{fig.nonint} 
(Color online) 
(Top) The non-interacting tight-binding band structure of 
the $c$-orbital (dashed line) and $f$-orbital (dash-dotted line).
Parameter values are $t_c = 1, \eps_c = 0, \eps_f = -1,$ and $t_f=-0.3$.
The horizontal solid line represents the chemical potential at half-filling.
(Bottom) The non-interacting Fermi surface at half-filling. 
The momentum path $(0,0) \rightarrow (\pi,\pi) \rightarrow  (\pi,0) \rightarrow (0,0)$ is also shown (thick straight line).}
\end{center}
\end{figure}

The non-interacting tight-binding band structure and 
corresponding Fermi surface at half-filling are shown in Fig.~\ref{fig.nonint}. 
At this parameter set, $c$- and $f$-band are overlapping each other. 
The level difference $\eps_f - \eps_c = -1$ causes an imbalance 
between the $c$- and $f$-electrons density, i.e., $n_c = 0.34$ and $n_f = 0.66$, 
where $n_c$ and  $n_f$ are the average density of $c$- and $f$-electrons respectively,
and we can see that 
$\mb{Q}=(\pi,\pi)$ is not a nesting vector of the Fermi surface. 
Thus, in this paper, we do not consider any periodic modulations characterized by $\mb{Q}$, such as the CDW phase,
which is realized with small energy difference $|\eps_c - \eps_f|$ and $n_c = n_f = 0.5$.~\cite{Farkasovsky} 
 
\subsection{Variational cluster approximation}

In order to analyze the EI state of EFKM, we apply the VCA.~\cite{Potthoff1}
Here we briefly review the formulation of the VCA in order to make our paper self-contained.
Following Potthoff,~\cite{Potthoff2} the grand potential functional is given as 
\begin{equation}
\Omega[ \bs{\Sigma} ] = {\cal F} [ \bs{\Sigma} ] - \mr{Tr} \ln \left( -\bs{G}_{0}^{-1} + \bs{\Sigma} \right)
\end{equation} 
where ${\cal F} [\bs{\Sigma}]$ is the Legendre transform of the Luttinger-Ward functional $\Phi[\bs{G}]$,~\cite{Luttinger} 
$\bs{G}_{0}$ is the non-interacting Green's function, 
and we call $\bs{\Sigma}$ the trial self-energy. 
$\mr{Tr}$ represents the sum over fermionic Matsubara frequencies with temperature $T$ and 
trace over the single-particle basis,
The explicit definition of $\mr{Tr}$ will be given later (in Eq.~(\ref{eq.Tr})).
The stationarity condition $\delta \Omega[\mb{\Sigma}]/\delta \bs{\Sigma} = 0$ gives the Dyson equation,
and the functional gives the grand potential of the system at the stationary point.~\cite{Potthoff2}

SFT~\cite{Potthoff2} provides a way to compute $\Omega$ by using the fact that 
the functional form of ${\cal F} [\bs{\Sigma}]$ depends only on 
the interaction terms of the Hamiltonian.
Here we introduce a so-called reference system, 
which consists of disconnected finite-size clusters forming a super-lattice. 
Note that, because the interaction term of the Hamiltonian Eq.~(\ref{eq.ham}) is local, 
the original and reference systems have the same interaction term. 
Therefore the functional form of ${\cal F}[\bs{\Sigma}]$ is unchanged.
The exact grand potential of the reference system is given as $\Omega' = {\cal F} [\bs{\Sigma}'] - \mr{Tr} \ln (-\bs{G}{'}_{0}^{-1} + \bs{\Sigma}')$, where
$\bs{\Sigma}'$ is the exact self-energy of the reference system.
Then, by restricting the trial $\bs{\Sigma}$ to $\bs{\Sigma}'$, we can omit the functional ${\cal F} [\bs{\Sigma}']$ and obtain 
\begin{equation}\label{eq.Omega}
\Omega [\bs{\Sigma}'] = \Omega' - \mr{Tr} \ln \left( \bs{I}-\bs{V}\bs{G}{'} \right) 
\end{equation}
where $\bs{I}$ is the unit matrix, 
$\bs{V}  \equiv  \bs{G}{'}_{0}^{-1} - \bs{G}_{0}^{-1}$ represents the difference of the one-body terms between the original and reference systems,
and $\bs{G}{'}=(\bs{G}{'}_{0}^{-1} - \bs{\Sigma}')^{-1}$ is the exact Green's function of the reference system. 
The size of these matrices are $2L_{\mr{c}} \times 2L_{\mr{c}}$, where $L_{\mr{c}}$ is the number of sites within a disconnected finite-size cluster.

The trial self-energy for the variational method is generated 
from the exact self-energy (or the exact Green's function) of the reference system.
The Hamiltonian of the reference system is defined as 
\begin{eqnarray}
&{\cal H}'            &= {\cal H} + {\cal H}_{\mr{pair}} + {\cal H}_{\mr{local}}, \\
&{\cal H}_{\mr{pair}} &= \Delta' \sum_{i} \left(c_{i}^\dag f_{i}  + \mr{H.c.} \right), \\
&{\cal H}_{\mr{local}}&= \eps'   \sum_{i} \left(n_{i c}           + n_{i f}   \right),
\end{eqnarray}
where the Weiss field for the on-site electron-hole pairing $\Delta'$ and 
the orbital-independent potential $\eps'$ are variational parameters, 
which are optimized based on the variational principle, i.e., 
$\left(\partial \Omega/\partial \Delta', \partial \Omega/\partial \eps' \right) = (0,0)$.  
Note that the solution with $\Delta' \not= 0$ corresponds to the spontaneous EI state.  
$\eps'$ is introduced in order to calculate the average particle density $n$ correctly.~\cite{Aichhorn}  
Then we solve the ground-state eigenvalue problem ${\cal H}'|\psi_0\rangle = E_0 |\psi_0\rangle $ 
of a finite-size cluster and calculate the trial single-particle Green's function 
by the Lanczos exact-diagonalization method. 
The Green's function matrix in Eq.~(\ref{eq.Omega}) is defined as  
\begin{eqnarray}
\bs{G}{'} (\w)=
\left( \begin{array}{cc}
\bs{G}{'}^{cc} (\w) & \bs{G}{'}^{cf} (\w) \\
\bs{G}{'}^{fc} (\w) & \bs{G}{'}^{ff} (\w)
\end{array} \right),
\end{eqnarray}
where  $\bs{G}^{\alpha \beta}$'s are the $L_{\mr{c}} \times L_{\mr{c}}$ matrices. 
Each matrix element is defined as
\begin{equation}
\begin{split}
G{'}^{\alpha \beta}_{ij} (\w) = 
\langle \psi_0 | \alpha_{i}       \frac{1}{\w - {\cal H}' + E_0} \beta_{j}^\dag | \psi_0 \rangle \\ +
\langle \psi_0 | \beta_{j}^\dag  \frac{1}{\w + {\cal H}' - E_0} \alpha_{i}      | \psi_0 \rangle
\end{split} 
\end{equation}
and these are calculated by the standard Lanczos technique. 
The matrix $\bs{V}$ is given as
\begin{eqnarray}
\bs{V} (\mb{K})= \left(
\begin{array}{cc}
 \bs{T}^{c} (\mb{K}) - \eps' \bs{I} & -\Delta' \bs{I} \\
-\Delta' \bs{I}                     &  \bs{T}^{f} (\mb{K}) - \eps' \bs{I}
\end{array} \right)
\end{eqnarray}
where $\bs{T}^{\alpha} (\mb{K})$ is the inter-cluster hopping matrix for $\alpha$-electrons. 
The matrix elements are given as 
$T_{ij}^{\alpha} (\mb{K}) = t_{\alpha} \sum_{\mb{X},x} \e^{\imag \mb{K} \cdot \mb{X}} \delta_{i+x,j} \delta_{\mb{R+X,R}'}$, 
where $x$ denotes the neighboring sites of the $i$-th site and 
$\mb{X}$ denotes the neighboring clusters of the $\mb{R}$-th cluster.  

Now all the physical quantities are diagonalized for Matsubara frequencies and super-lattice wave vectors, 
but not for orbitals and sites within a cluster.
Thus $\mr{Tr}$ for a quantity $\bs{A}$ is written explicitly as
\begin{equation}\label{eq.Tr}
\mr{Tr} \bs{A} = T \sum_{\w_n} \e^{\imag \w_n 0+} \sum_{\mb{K}} \sum_{\alpha=c,f} \sum_{i=1}^{L_{\mr{c}}} A_{ii}^{\alpha \alpha}(\mb{K},\imag \w_n). 
\end{equation}
The $\mb{K}$-summation is done in the reduced Brillouin zone of the superlattice. 
For numerical calculations of $\Omega$, 
the Matsubara-frequency sum is transformed 
to a contour integral with complex Fermi function $f(\w)=1/(\e^{\w/T}+1)$ by the theorem of residuum.
Then the contour is deformed to a path enclosing the real axis by use of the convergence factor. 
Finally we obtain an expression for the functional,  
\begin{equation}
\Omega = \Omega' - \oint \frac{\dd \w}{2 \pi \imag} \sum_{\mb{K}} \mathrm{ln} \mathrm{det} \left(\bs{I}-\bs{V}(\mb{K}) \bs{G}'(\w) \right).
\end{equation}

The single-particle Green's functions are calculated by CPT~\cite{Senechal2} 
with the optimized variational parameters. 
The CPT Green's function is defined as 
\begin{equation}
{\cal G}^{\alpha \beta} (\mb{k},\w) = \frac{1}{L_{\mr{c}}} \sum_{i,j=1}^{L_{\mr{c}}} G_{\mr{CPT},ij}^{\alpha \beta}(\mb{k},\w) \e^{-\imag \mb{k} \cdot (\mb{r}_i-\mb{r}_j)}
\end{equation}
where $\mb{r}_i$ is the position of the $i$-th site within a disconnected finite-size cluster and  
$\bs{G}_{\mr{CPT}}(\mb{k},\w) = \bs{G}{'}(\w) (\bs{I} -  \bs{V}(\mb{k}) \bs{G}{'}(\w))^{-1}$.
The wave-vector $\mb{k}$ can take arbitrary values in the 1$^\mr{st}$ Brillouin-zone.
Here we define ${\cal G}^{cc} (\mb{k},\w)$, ${\cal G}^{ff} (\mb{k},\w)$, and ${\cal G}^{cf} (\mb{k},\w)$
as the single particle $c$-electron, $f$-electron, and anomalous Green's functions, respectively. 

A cluster of the size $L_{\mr{c}}=8$ (16-orbital) is used as a reference system, 
thus the effects of statical and dynamical electron correlation within the cluster size are taken into account.  
Details of VCA can be found in Refs.~[\onlinecite{Potthoffreview, Senechalreview}].


\subsection{Hartree-Fock approximation}

It was reported that the ground-state phase diagram of the two-dimensional EFKM obtained by HFA 
quantitatively agrees with that by CPMC~\cite{Batista1} in the weak-to-intermediate coupling regime.~\cite{Farkasovsky}
To compare VCA results with HFA results, we briefly review the mean-field theory for EI state of this model.~\cite{Ihle} 
Applying the HFA to the interaction term in the original Hamiltonian Eq.(\ref{eq.ham}), i.e., 
$        c_i^\dag c_i         f_i^\dag f_i  \rightarrow  
 \langle c_i^\dag c_i \rangle f_i^\dag f_i 
+\langle f_i^\dag f_i \rangle c_i^\dag c_i         
-\langle f_i^\dag c_i \rangle c_i^\dag f_i 
-\langle c_i^\dag f_i \rangle f_i^\dag c_i, $
where $\bra \cdots \ket$ denotes the ground-state expectation value,
and switching to momentum space, we obtain the mean-field Hamiltonian
\begin{eqnarray} \label{eq.Hmf}
&&{\cal H}_{\rm HFA}=\sum_{\mb{k}} 
\left(
\begin{array}{cc}
 c_{\mb{k}}^{\dag}
&f_{\mb{k}}^{\dag}
\end{array}\right)
\left(
\begin{array}{cc}
\eps_{\mb{k}c} & -\Delta \\
-\Delta        & \eps_{\mb{k}f}
\end{array}
\right) 
\left(
\begin{array}{c}
 c_{\mb{k}} \\
 f_{\mb{k}}
\end{array}
\right), \notag \\
&&\eps_{\mb{k} \alpha} = 2t_\alpha (\cos{k_x} + \cos{k_y}) + \eps_{\alpha} - \mu + U n_{\bar{\alpha}}, \notag \\
&& n_\alpha   = \frac{1}{L} \sum_{\mb{k}} \langle \alpha_{\mb{k}}^\dag \alpha_{\mb{k}} \rangle, \notag \\ 
&&\Delta = \frac{U}{L} \sum_{\mb{k}} \langle f^\dag_{\mb{k}} c_{\mb{k}} \rangle,
\label{orderHFA}
\end{eqnarray}
where $L$ is the number of lattice sites and $c_{\mb{k}}$ ($f_{\mb{k}}$) is the Fourier transform of $c_{i}$ ($f_{i}$). 
$\alpha=c,f$ represents the orbital index and $\bar{\alpha}$ denotes the other orbital of $\alpha$, i.e., $\bar{c} = f$ and vice versa.
The order parameter $\Delta$ describes the coherent exciton formation between $c$-electrons and $f$-holes.
Here we assumed $\Delta$ is real without loss of generality. Introducing the fermionic quasi-particles defined as  
\begin{equation}
\left(
\begin{array}{c}
\gamma_{\mb{k}}^+\\
\gamma_{\mb{k}}^-\\
\end{array}
\right)=\left(
\begin{array}{cc}
u_{\mb{k}} &  v_{\mb{k}} \\
v_{\mb{k}} & -u_{\mb{k}}
\end{array}
\right)\left(
\begin{array}{c}
c_{\mb{k}}\\
f_{\mb{k}}\\
\end{array}
\right)
\end{equation}
with $u_{\mb{k}}^2 + v_{\mb{k}}^2 = 1$ and diagonalizing the matrix in Eq. (\ref{eq.Hmf}) for each $\mb{k}$, we obtain the mean-field Hamiltonian  
\begin{equation}
{\cal H}_{\mr{HFA}} = \sum_{\mb{k}} \left( E^+_{\mb{k}} \gamma_{\mb{k}}{^+}^{\dag} \gamma_{\mb{k}}{^+}
                                         + E^-_{\mb{k}} \gamma_{\mb{k}}{^-}^{\dag} \gamma_{\mb{k}}{^-} \right),
\end{equation}
with the quasi-particle dispersion
\begin{eqnarray}
&& E^{\pm}_{\mb{k}} = \frac{1}{2} \left(\eps_{\mb{k}c} + \eps_{\mb{k}f} \right) \pm \sqrt{\xi_{\mb{k}}^2 + \Delta^2}, \label{eq.Ek} \\
&& \xi_{\mb{k}} = \frac{1}{2} (\eps_{\mb{k}c} - \eps_{\mb{k}f}).
\end{eqnarray}
Self-consistency equations for the particle density and the order parameter are
\begin{eqnarray}
&& n_c    = \frac{1}{L} \sum_{\mb{k}} \left( u_{\mb{k}}^2 f(E_{\mb{k}}^{+}) + v_{\mb{k}}^2 f(E_{\mb{k}}^{-}) \right), \\
&& n_f    = \frac{1}{L} \sum_{\mb{k}} \left( v_{\mb{k}}^2 f(E_{\mb{k}}^{+}) + u_{\mb{k}}^2 f(E_{\mb{k}}^{-}) \right), \\
&& \Delta = \frac{U}{L} \sum_{\mb{k}} u_{\mb{k}} v_{\mb{k}} \left( f(E_{\mb{k}}^{+}) - f(E_{\mb{k}}^{-})  \right), 
\end{eqnarray}
respectively, where the quasi-particle density $\langle \gamma^{\pm \dag}_{\mb{k}} \gamma^{\pm}_{\mb{k}} \rangle$ 
is replaced by the Fermi function $ f(E^{\pm}_{\mb{k}}) = 1/(\e^{E^{\pm}_{\mb{k}}/T} + 1)$. 
The coefficients are given as
\begin{eqnarray}
&& u_{\mb{k}}^2 = \frac{1}{2} \left( 1 + \frac{\xi_{\mb{k}}}{ \sqrt{\xi_{\mb{k}}^2 + \Delta^2 }} \right), \\
&& v_{\mb{k}}^2 = \frac{1}{2} \left( 1 - \frac{\xi_{\mb{k}}}{ \sqrt{\xi_{\mb{k}}^2 + \Delta^2 }} \right), \\
&& u_{\mb{k}} v_{\mb{k}} = -\frac{\Delta}{2\sqrt{\xi_{\mb{k}}^2 + \Delta^2}} \label{eq.ukvk}.
\end{eqnarray}
The parameters $n_c$, $n_f$, and $\Delta$ are determined by solving the above equations self-consistently.

Inverting the matrix $(\w - {\cal H}_{\mb{k}})$, where ${\cal H}_{\mb{k}}$ is the $2 \times 2$ matrix in the mean-field Hamiltonian (\ref{eq.Hmf}), 
we obtain the $c$-orbital, $f$-orbital, and anomalous Green's function 
\begin{eqnarray} \label{eq.HFAGF}
&&{\cal G}^{cc}_{\mr{HFA}}(\mb{k},\w) = \frac{1}{\w - \eps_{\mb{k}c} - \frac{\Delta^2}{\w - \eps_{\mb{k}f}}}, \\
&&{\cal G}^{ff}_{\mr{HFA}}(\mb{k},\w) = \frac{1}{\w - \eps_{\mb{k}f} - \frac{\Delta^2}{\w - \eps_{\mb{k}c}}}, \\
&&{\cal G}^{cf}_{\mr{HFA}}(\mb{k},\w) = \frac{\Delta} { (\w - \eps_{\mb{k}c} ) (\w - \eps_{\mb{k}f})  - \Delta^2}, 
\end{eqnarray}
respectively. From the imaginary part of each of the Green's functions, we obtain the $c$-orbital, $f$-orbital, and anomalous spectral function 
\begin{eqnarray}
&&A^c_{\mr{HFA}}(\mb{k},\w) = u_{\mb{k}}^2                 \delta(\w - E^+_\mb{k}) + v_{\mb{k}}^2 \delta(\w - E^-_\mb{k}),   \\
&&A^f_{\mr{HFA}}(\mb{k},\w) = v_{\mb{k}}^2                 \delta(\w - E^+_\mb{k}) + u_{\mb{k}}^2 \delta(\w - E^-_\mb{k}),   \\
&&  F_{\mr{HFA}}(\mb{k},\w) = u_{\mb{k}} v_{\mb{k}} \left (\delta(\w - E^+_\mb{k}) -              \delta(\w - E^-_\mb{k}) \right),   
\end{eqnarray}
respectively.


\section{Results of calculation}
\label{sec_result}



\subsection{Order parameter and single-particle gap}


\begin{figure}[t]
\begin{center}
\includegraphics[width=17pc]{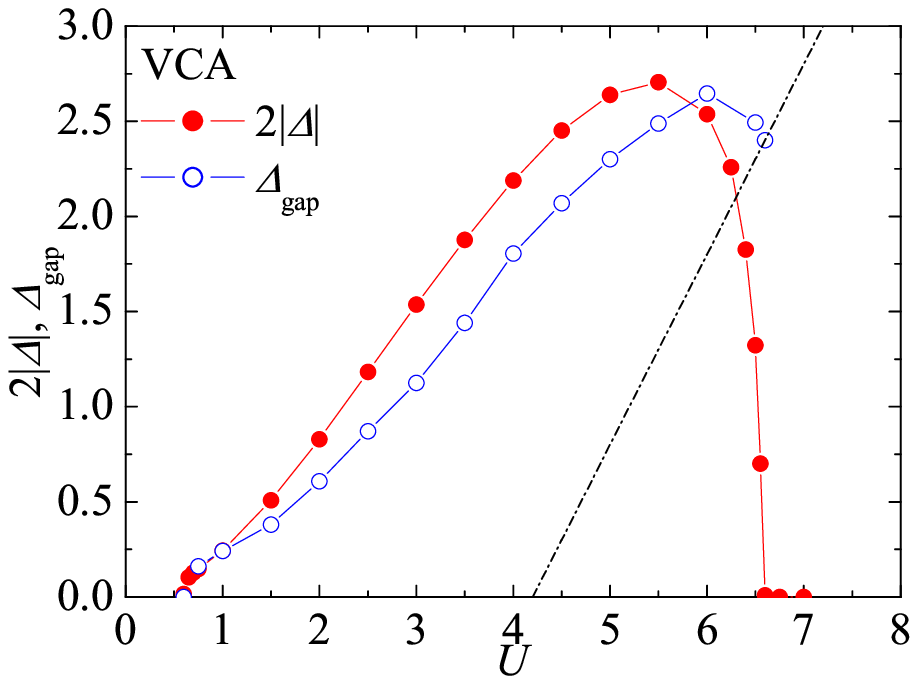} \\
\includegraphics[width=17pc]{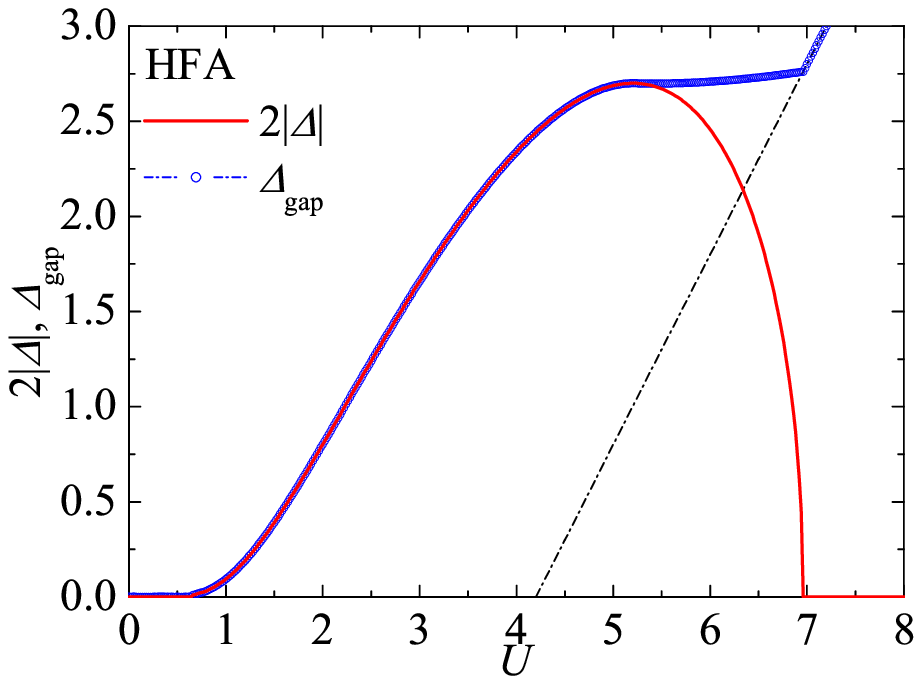}
\caption{
\label{fig.order} 
(Color online)
$U$ dependence of the order parameter $2|\Delta|$ and the single-particle gap $\Delta_{\mr{gap}}$ calculated by VCA (upper panel).  
HFA results for  $2|\Delta|$ and $\Delta_{\mr{gap}}$ are also shown (lower panel). 
The dash-dotted straight line indicates the single-particle gap in the normal ($\Delta=0$) state.}
\end{center}
\end{figure}

We first calculate the $U$ dependence of the order parameter for exciton condensation 
\begin{equation}
2\Delta = \frac{U}{L} \sum_{i}  \langle c_i^\dag f_i + \mr{H.c.} \rangle 
\end{equation}
and the single-particle excitation gap defined as
\begin{equation}
\Delta_{\mr{gap}} = \mu^+ - \mu^-
\end{equation} 
where $\mu^{+(-)}$ is the upper (lower) bound of the chemical potential. 
Calculated results by VCA (HFA) are shown in the upper (lower) panel of the Fig.~\ref{fig.order}.
The factor 2 for the order parameter is introduced in order to compare with the single-particle gap, by analogy with the BCS mean-field theory.

We can see from the results that there are 
not only a lower bound of the Coulomb interaction strength $U_{\mr{c}1}$ 
but also an upper bound  $U_{\mr{c}2}$ for the EI state. 
The obtained values are $(U_{\mr{c}1}, U_{\mr{c}2})$ = (0.65, 6.6) for VCA and $(U_{\mr{c}1}, U_{\mr{c}2})$ = (0.66, 6.95) for HFA, respectively.
The existence of the upper bound $U_{\mr{c}2}$ seems to contradict 
to the case of the attractive Hubbard model, which has no $U_{\mr{c}2}$.~\cite{Toschi,Koga}
What happened at $U=U_{\mr{c}2}$ is that, the Hartree potential makes the $f$-band fully occupied and $c$-band empty,  
so that there is no Coulomb interaction between $c$-electrons and $f$-holes. 
Thus the system is simply a band insulator above $U_{\mr{c}2}$.

Note that, which band becomes empty or full at large $U$ is determined 
by the particle density of each orbital at $U=0$.
In our case, $n_c < n_f$ at $U$=0.
Therefore the Hartree potential for $c$-band is larger than that for $f$-band. 
Thus, with increasing $U$, the $c$-band is pushed up rather than $f$ band is, and finally $c$-band becomes empty. 

In the weak-to-intermediate coupling regime ($U \lesssim 5$), 
both the order parameter and single-particle gap increase with increasing $U$
with the relation $2|\Delta| \simeq \Delta_{\mr{gap}}$ (upper panel). 
This result is consistent with the relation $2|\Delta| = \Delta_{\mr{gap}}$ from HFA (lower panel).     
In the strong-coupling regime ($U \gtrsim 5$), the order parameter rapidly decreases with increasing $U$ but the single-particle gap remains open.   
If we can assume that the energy scale of the single-particle gap $\Delta_\mr{gap}$ and 
order parameter $2|\Delta|$ may correspond to that of the characteristic temperature for the exciton 
formation ($T_{\mr{ex}}$) and critical temperature for the condensation of excitons ($T_{\mr{c}}$), 
respectively, then the two temperatures should be comparable ($T_{\mr{ex}} \simeq T_{\mr{c}}$) in 
the weak-coupling (BCS) regime but may be well separated ($T_{\mr{ex}} \gg T_{\mr{c}}$) in the strong-coupling (BEC) regime.  
The BCS-BEC crossover may then be expected in this model although our calculations are done at zero temperature.


\subsection{Momentum distribution function}

\begin{figure}[htb]
\begin{center}
\includegraphics[width=19pc]{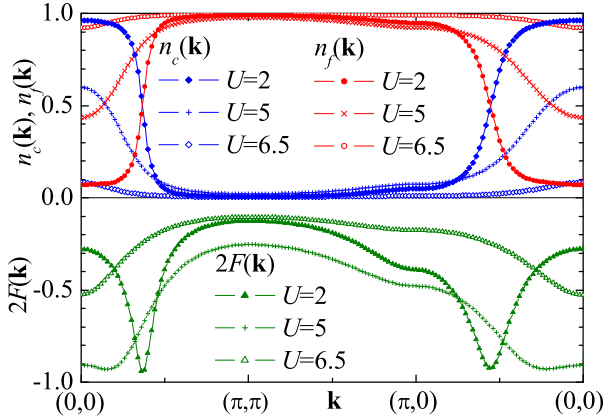}
\caption{
\label{fig.nk} (Color online) Calculated momentum distribution functions 
$n_c(\mb{k})$, $n_f(\mb{k})$, and $F(\mb{k})$ at $U=2$, $U=5$, and $U=6.5$.}
\end{center}
\end{figure}

We then consider the $c$-electron, $f$-electron, and anomalous momentum distribution functions defined as 
\begin{eqnarray}
n_\alpha(\mb{k}) &=& \oint_{C_{<}} \frac{\dd z}{2\pi \imag} {\cal G}^{\alpha \alpha}(\mb{k},z), \\
F(\mb{k})   &=& \oint_{C_{<}} \frac{\dd z}{2\pi \imag} {\cal G}^{cf}(\mb{k},z),
\end{eqnarray}
respectively, where the contour integral path $C_<$ encloses the poles of the integrand on the real axis below the chemical potential.
The results are shown in Fig.~\ref{fig.nk}. 
Here we define, by analogy with the BCS mean-field theory, the ``Fermi momentum" $\mb{k}_{\mr{F}}$ as 
\begin{equation} \label{kF}
n_c(\mb{k}_{\mr{F}}) = n_f(\mb{k}_{\mr{F}}) = 0.5.
\end{equation}
Actually in HFA, this definition means that the single particle gap is identical to $2|\Delta|$ and $2|F(\mb{k})|=1$ at $\mb{k}_{\mr{F}}$
(see Eq.~(\ref{eq.Ek}) and Eq.~(\ref{eq.ukvk}) respectively). 
At $U=2$, $F(\mb{k})$ shows the sharp peak at $\mb{k}_{\mr{F}}$, indicating the existence of weakly bound electron-hole pairs.  
At $U=6.5$, $\mb{k}$-dependence of the peak intensity of the anomalous Green's function is weak and $F(\mb{k})$ is spread out in momentum space.  
Thus, in real space, small electron-hole pairs exist in the strong-coupling regime.  

\begin{figure}[ht]
\begin{center}
\includegraphics[width=19pc]{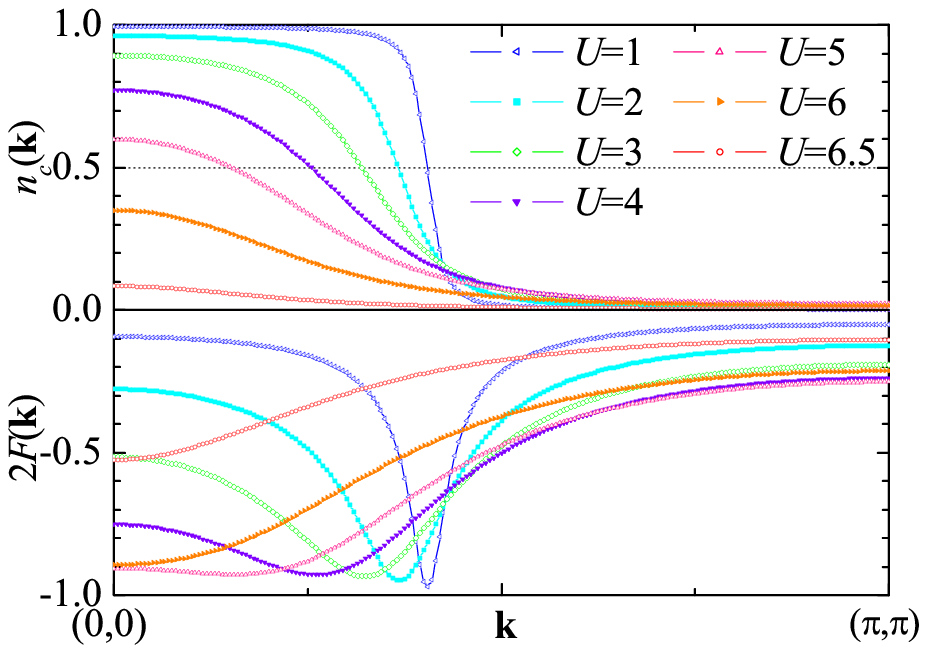}\\
\includegraphics[width=19pc]{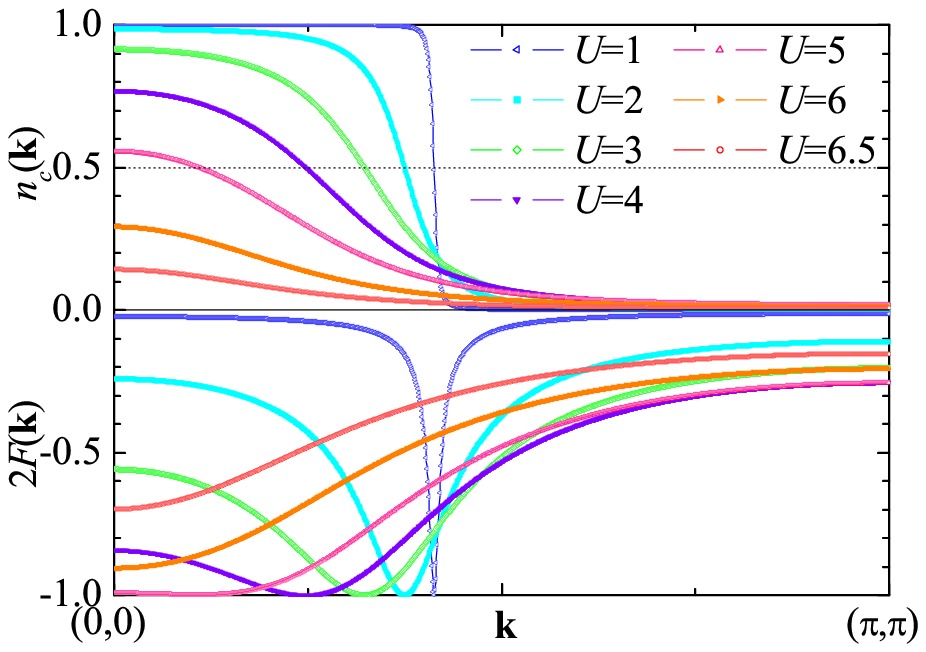}
\caption{
\label{fig.nk00pp} 
(Color online) The momentum distribution functions $n_c(\mb{k})$ and $2F(\mb{k})$ 
along $(0,0) \rightarrow (\pi,\pi)$ for various $U$ calculated by VCA (upper panel) and HFA (lower panel).
The $f$-electron momentum distribution functions $n_f(\mb{k}) = 1 - n_c(\mb{k})$ are not shown.}
\end{center}
\end{figure}

To see the  $U$ dependence of the momentum distribution functions in more detail, 
we show the results along the $(0,0) \rightarrow (\pi,\pi)$ line in Fig.~\ref{fig.nk00pp}.
We can see that, in the weak-coupling regime, 
$n_c(\mb{k})$ drops sharply across $\mb{k}_{\mr{F}}$ and $F(\mb{k})$ is peaked at $\mb{k}_{\mr{F}}$.
With increasing $U$, $\mb{k}_{\mr{F}}$ approaches $(0,0)$ 
because the Hartree potential for the $c$-electron reduces the $c$-electron density,
and $F(\mb{k})$ becomes broad in momentum space, 
indicating that the radius of electron-hole pairs becomes small in real space. 
When $U$ reaches the crossover regime ($U \sim 5$), 
we have no $\mb{k}_\mr{F}$ and $|F(\mb{k})|$ decreases for all momenta with increasing $U$.
This behavior is consistent with the rapid decrease of $|\Delta|$ in the strong-coupling regime (see Fig.~\ref{fig.order}).

\subsection{Coherence length}

\begin{figure}[ht]
\begin{center}
\includegraphics[width=20pc]{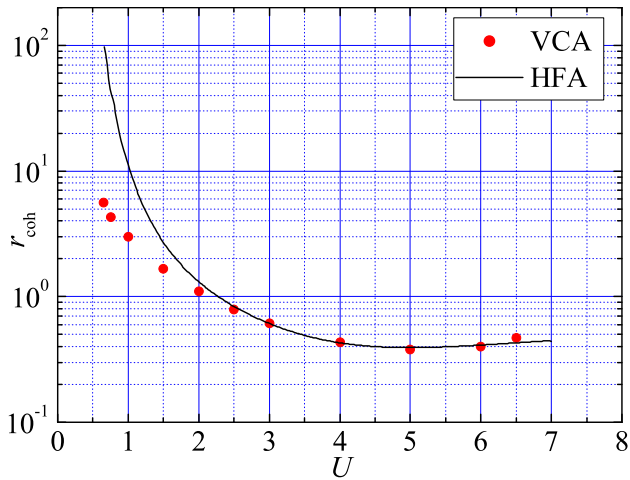}
\caption{\label{rcoh} (Color online) 
$U$ dependence of the coherence length calculated by VCA (circles) and HFA (solid line). }
\end{center}
\end{figure}

We also evaluated the coherence length defined as
\begin{equation} \label{eq.rcoh}
r_{\mr{coh}} = \sqrt{\frac{\sum_{\mb{k}} |\nabla_{\mb{k}} F(\mb{k})|^2}{ \sum_{\mb{k}} |F(\mb{k})|^2}}
\end{equation}
in order to see the spatial coherence of the excitons directly.  
The $\mb{k}$-summations were done with $100 \times 100$ $\mb{k}$-points
in the 1$^{\mr{st}}$ Brillouin zone.
For VCA calculations, the derivative with respect to $k_{x,y}$ was evaluated 
by the 4-point finite difference, while for HFA calculations, 
the analytical expression for $\nabla_{\mb{k}} F(\mb{k})$ was used.  
Calculated results are shown in Fig.~\ref{rcoh}.  
In the weak-coupling regime, $r_{\mr{coh}}$ is spread widely, 
about several lattice-constants and rapidly decreases with increasing $U$. 
Note that the calculated result of $r_{\mr{coh}}$ by VCA is considerably smaller 
than that by HFA in the weak-coupling regime, especially for $r_{\mr{coh}} > 1$.
At $U=2 \sim 3$, $r_{\mr{coh}}$ is already the size of the lattice-constant.
A similar rapid decrease of the coherence length of the Cooper pairs as a function of the Coulomb interaction strength 
was also reported in a detailed exact-diagonalization study on the attractive Hubbard model.~\cite{Ohta}
Furthermore, we find that $r_{\mr{coh}}$ has a shallow minimum at $U \simeq 5$ where the system is in crossover regime.
This is because the denominator $\sum_{\mb{k}} |F(\mb{k})|^2 $ in Eq.(\ref{eq.rcoh}) is largest in the crossover regime (see Fig.~\ref{fig.nk00pp}). 
Then $r_{\mr{coh}}$ slightly increases again with increasing $U$.


\subsection{Single-particle spectra and density of state}

\begin{figure*}[ht]
\begin{center}
\includegraphics[width=38pc]{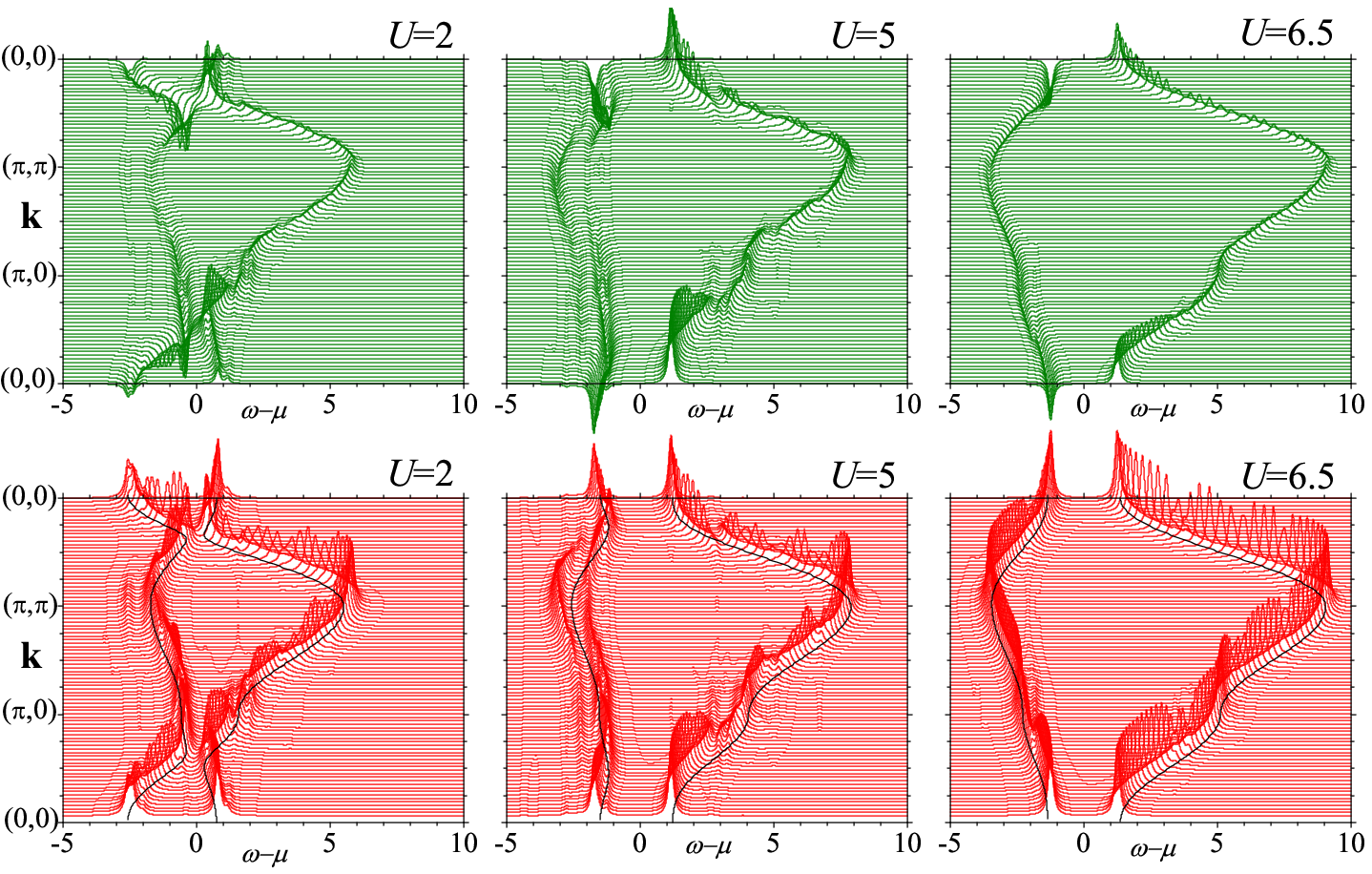}
\caption{\label{Akw} (Color online) Calculated 
anomalous spectra $F(\mb{k},\w)$ (upper panel) and 
single-particle excitation spectra $A(\mb{k},\w)$ (lower panel) at 
$U=2$ (left),
$U=5$ (center), and 
$U=6.5$ (right).
In the lower panel, the HFA quasi-particle dispersion $E^{\pm}_{\mb{k}}$ is also shown (solid line).
The artificial Lorenzian broadening $\eta=0.1$ is used.}
\end{center}
\end{figure*}

\begin{figure}[ht]
\begin{center}
\includegraphics[width=18pc]{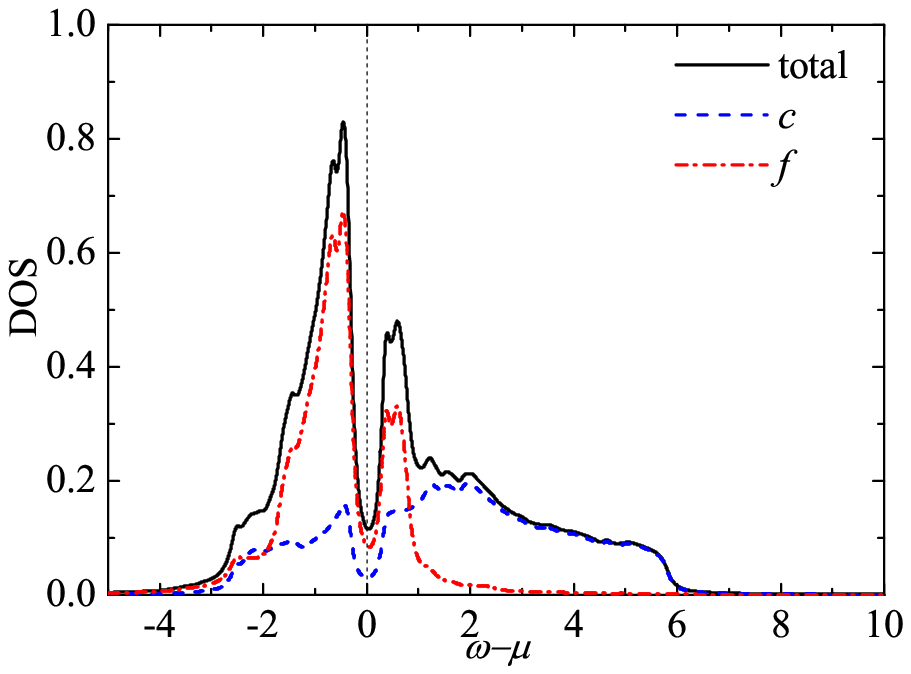} \\
\includegraphics[width=18pc]{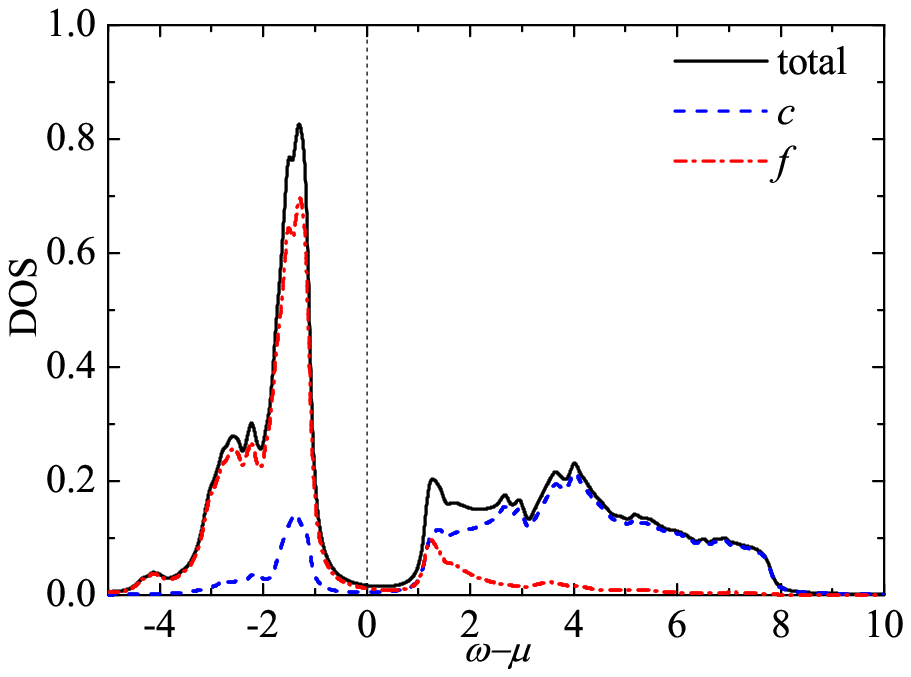} \\
\includegraphics[width=18pc]{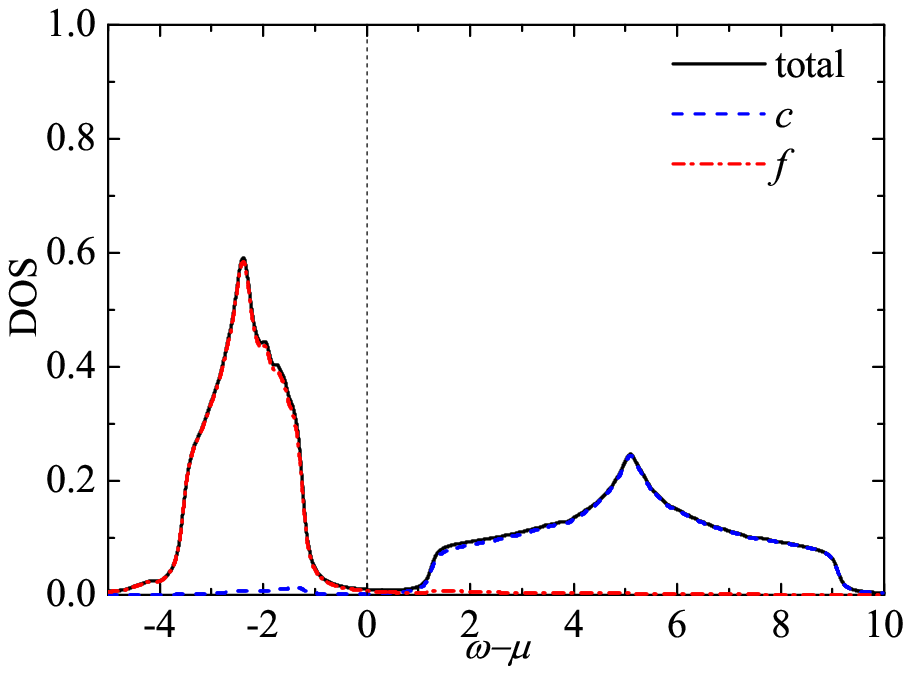} 
\caption{\label{DOS} (Color online) Calculated density of states at 
$U=2$ (top),
$U=5$ (middle), and 
$U=6.5$ (bottom).
The vertical line represents the Fermi energy.
The artificial Lorenzian broadening $\eta=0.1$ is used.}
\end{center}
\end{figure}

We also calculated the single-particle and anomalous excitation spectra 
\begin{eqnarray}
&& A (\mb{k},\w) = -\frac{1}{\pi} \Im \sum_{\alpha=c,f} {\cal G}^{\alpha \alpha} (\mb{k},\w+\imag \eta) \\
&& F   (\mb{k},\w) = -\frac{1}{\pi} \Im {\cal G}^{cf} (\mb{k},\w+\imag \eta)
\end{eqnarray} 
and the density of states (DOSs)
\begin{equation}
\rho^{\alpha} (\w) = \frac{1}{L} \sum_{\mb{k}} A^{\alpha}(\mb{k},\w) 
\end{equation}
at $U=2$ (BCS regime), $U=5$ (Crossover regime), and $U=6.5$ (BEC regime).
The results are shown in Fig.~\ref{Akw} and Fig.~\ref{DOS}, respectively.
The HFA quasi-particle dispersion $E_{\mb{k}}^{\pm}$ is also shown in the lower panel of Fig.~\ref{Akw}. 
Note that, because of the artificial supercell structure introduced by the VCA, 
the spectra show artificial gaps due to Brillouin zone folding.
At $U=2$, $A(\mb{k},\w)$ shows a small gap at the Fermi momentum $\mb{k}_\mr{F}$ defined in Eq.~(\ref{kF}).  
$F(\mb{k},\w)$ shows sharp peak near $\mb{k}_\mr{F}$ and its intensity rapidly decreases as the 
momentum goes away from $\mb{k}_\mr{F}$ or the frequency goes away from the Fermi level $\mu$.  
At $U=5$, the incoherent continua appear in the spectral function.
We can see from the DOS and anomalous Green's functions that 
both the single particle gap and hybridization are large. 
At $U=6.5$, $A(\mb{k},\w)$ shows a semiconductor-like dispersion mainly due to the Hartree potential.  
The momentum dependence of the intensity of $F(\mb{k},\w)$ is weaker than that at $U=2$ and $5$.  
Note that, although $U$ is large, the incoherent part of both $A(\mb{k},\w)$ and $F(\mb{k},\w)$ is smaller than 
that at $U=5$. The dispersion relation is well described by the HFA quasi-particle dispersion $E_{\mb{k}}^{\pm}$
both in the weak- and strong-coupling regime.
The reason will be discussed from the view point of the self-energy at Sec.~\ref{sec_discussion}.


\section{Discussions}
\label{sec_discussion}


\subsection{Why HFA works well}

Here we consider, from the view point of the self-energy, 
why the HFA is successful not only in the weak-coupling regime but also in the strong coupling regime for EFKM.
For simplicity, we neglect the order parameter $\Delta$ and Weiss field for electron-hole pairs $\Delta'$. 

\begin{figure}[ht]
\begin{center}
\includegraphics[width=20pc]{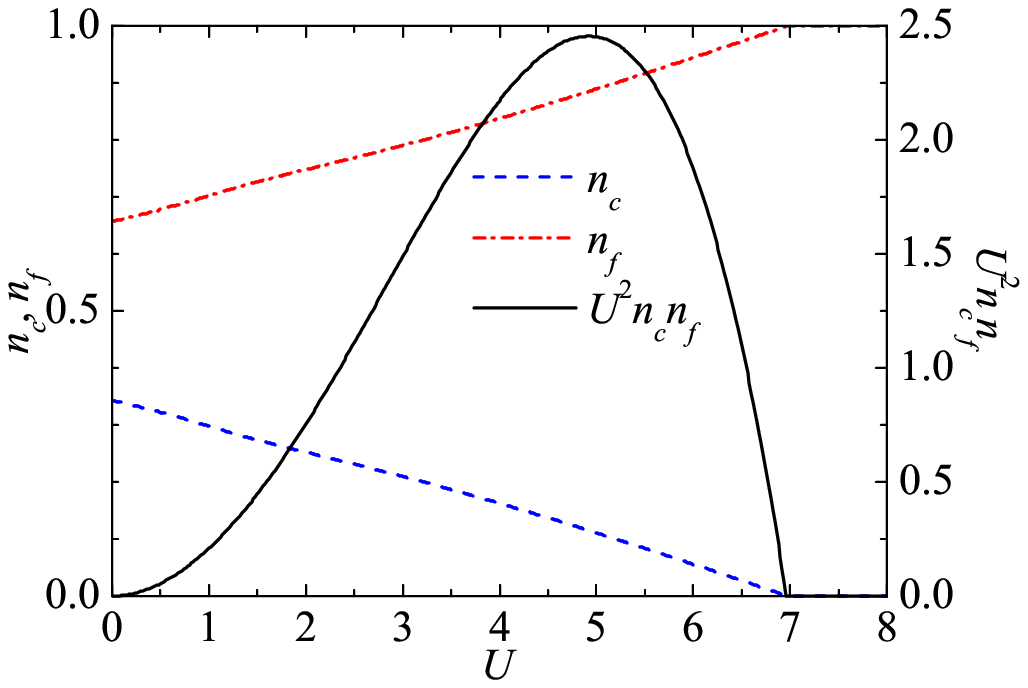}
\caption{\label{Usqncnf} (Color online) $U$ dependence of 
$n_c$ (dashed line), 
$n_f$ (dash-dotted line), and 
$U^2 n_c n_f$ (solid line) calculated by HFA.}
\end{center}
\end{figure}

Using the spectral representation,~\cite{Luttinger2} the self-energy can be written as
\begin{equation}\label{Sspect}
\Sigma(\bf{k},\w) ={\it g}_{\bf{k}} + \sum_{\nu} \frac{\sigma_{\bf{k},\nu}}{\w - \zeta_{\bf{k},\nu}},
\end{equation}
where 
$g_{\bf{k}}$ is the Hartree potential,~\cite{Eder} 
$\zeta_{\bf{k},\nu}$ is the $\nu$-th pole of the self-energy, and 
$\sigma_{\bf{k},\nu}$ is the corresponding spectral weight. 
In HFA, the Hartree potential is taken into account but the second term (frequency dependence of the self-energy) is neglected. 
Note that EFKM defined in Eq.~(\ref{eq.ham}) is nothing but the asymmetric Hubbard model. 
Therefore we can apply the sum-rule for the self-energy of the Hubbard model (see Ref.~[\onlinecite{Turkowski}] and Appendix).
Then total weight of the self-energy `neglected' in the HFA is
\begin{equation} \label{sumrule}
-\frac{1}{\pi} 
\lim_{\eta \rightarrow 0+} 
\int_{-\infty}^{\infty} \dd \w \Im \Sigma(\mb{k},\w+\imag \eta) 
=
\sum_{\nu} \sigma_{\mb{k},\nu} 
= U^2 n_c n_f.
\end{equation}
Here the half-filling condition $n_c + n_f=1$ is used.
The $U$ dependence of $n_c$, $n_f$, and $U^2 n_c n_f$ calculated by the HFA is shown in Fig.~\ref{Usqncnf}. 
We can see from the result that, with increasing $U$, the Hartree potential causes particle number imbalance of the $c$- and $f$-orbitals,
and makes $n_c n_f $ smaller.
Thus $U^2 n_c n_f $ has a maximum and decrease again with increasing $U$.

Moreover, we can see from the HFA Green's functions Eq.~(\ref{eq.HFAGF})
that, if the order parameter $\Delta$ is finite,  
the single-particle gap can be generated from  
the hybridization gap in the weak-coupling regime.
Thus HFA works well on the EFKM both in the weak- and strong-coupling regime.

The crucial differences between VCA and HFA will appear in one- and two-dimensional system at finite temperature, 
where the spontaneous symmetry breaking is absent.~\cite{MerminWagner,Hohenberg} 
Actually, it was reported that the critical temperature for exciton condensation of the EFKM evaluated by the SB technique
is lower than that by the mean-field due to the effects of electron correlations.~\cite{Zenker2}
Moreover, the effects of the spatial fluctuations, which are also completely neglected in the mean-field theory,
will also tend to destroy the ordering.


\subsection{Experimental implications}

\begin{figure}[ht]
\begin{center}
\includegraphics[width=20.5pc]{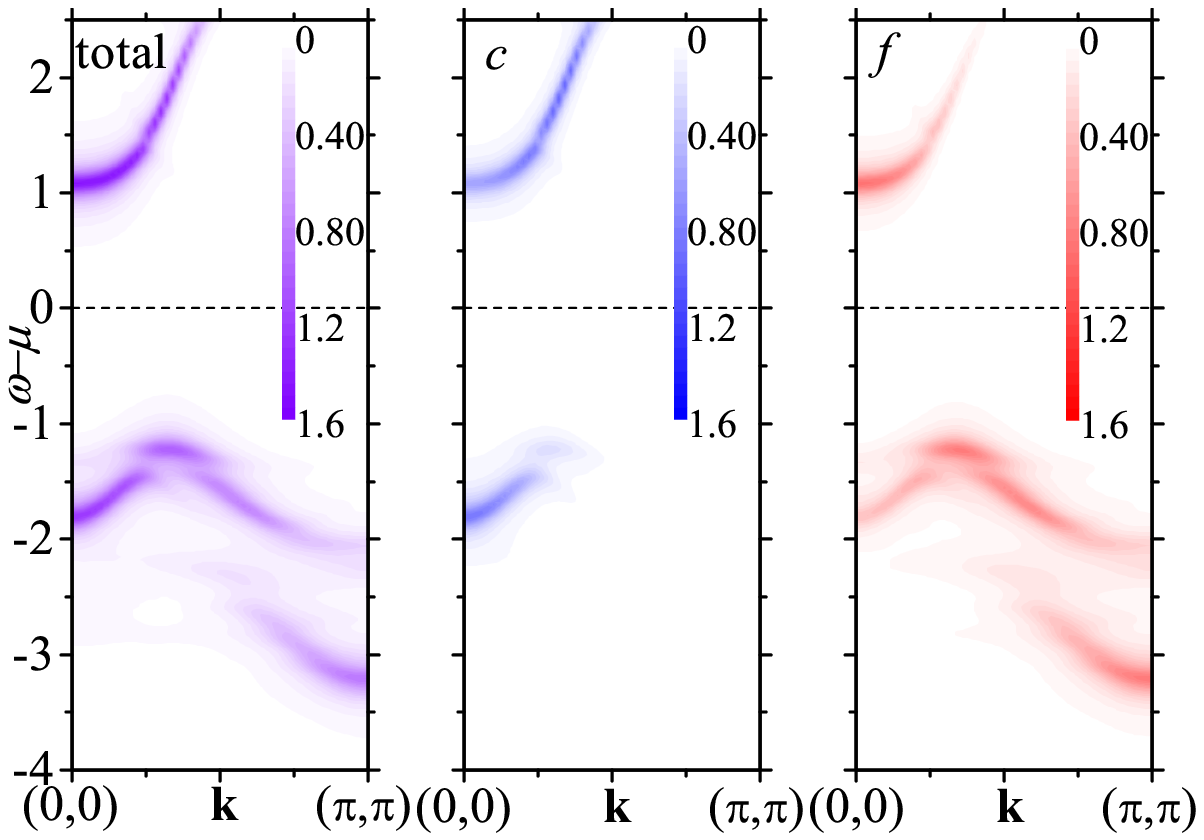}
\caption{\label{Akwcontour} (Color online) Calculated 
single-particle excitation spectra $A(\mb{k},\w),A^c(\mb{k},\w),$ and $A^f(\mb{k},\w)$ (from left to right) at $U=5$.
The horizontal line represents the Fermi energy.
The artificial Lorenzian broadening $\eta=0.1$ is used.}
\end{center}
\end{figure}

Since the EI order parameter is not necessarily identical 
with the single-particle gap, especially in BEC regime,    
an experimental evidence for the realization of EI should be 
signaled as the spontaneous hybridization between the valence and conduction bands.
Generally, ARPES experiments observe the imaginary part of the single-particle Green's function 
filtered by the dipole matrix element and the Fermi function.~\cite{Damascelli} 
The matrix element effects are determined by the selection rule from symmetries,
photon energy dependence of the cross section, etc.
Consideration about the matrix element effects becomes important 
for observation of Zhang-Rice singlet states in cuprate by ARPES.~\cite{Katagiri}
If one can resolve dominant orbitals for the valence and conduction band 
by use of matrix element effects, 
the spontaneous hybridization between 
the valence and conduction bands can be observed by ARPES
as the difference in the spectral intensity 
between above and below the critical temperature.
That is, the spectral intensity from the dominant conduction-band orbitals 
will be transferred to the valence-band top below the critical temperature.  
Therefore, experimental analyses of temperature dependence 
of the hybridization between the valence and conduction bands are desired. 

The recent ARPES measurements on quasi-one-dimensional Ta$_2$NiSe$_5$ were done 
with the photon energies $h\nu=10$ eV and $h \nu=23$ eV.~\cite{Wakisaka}
The experimental results of the energy distribution curve (EDC) at $T=40$ K showed that 
the EDC intensity is large near $\Gamma$ with $h \nu = 23$ eV, 
but small with $h \nu = 10$ eV.
From the cross section table of atoms,~\cite{Yeh}
Ta $5d$ weight should be large for $h\nu=23$ eV, while 
Se $4p$ weight should be large for $h\nu=10$ eV.
The result implies that 
the spectral weight of the conduction-band Ta $5d$ orbitals is transferred 
to the valence top due to the hybridization.
The similar spectral-weight transfer can be seen in our calculated result for 
the orbital-resolved excitation spectra shown in Fig.~\ref{Akwcontour}.
We can see from the results that, near $\mb{k}=(0,0)$, the spectral weight of 
the dominant conduction-band orbital ($c$-orbital) is transfered below the Fermi energy.
This spectral-weight transfer also can be seen in the calculated DOS at $U=5$ (middle panel of Fig.~\ref{DOS}). 
Thus, the temperature-dependent photoemission spectroscopy measurements 
with various photon energies are desired to identify the orbital character of the band structure and 
estimate the hybridization between the valence and conduction bands in Ta$_2$NiSe$_5$.  


\section{Summary}
\label{sec_summary}


In this paper we have analyzed the excitonic insulator (EI) state of the extended Falicov-Kimball model (EFKM) 
by using the variational cluster approximation (VCA) and Hartree-Fock approximation (HFA) at zero temperature.
We have calculated 
the EI order parameter, 
single-particle gap, 
momentum distribution functions, 
coherence length, and 
single-particle Green's functions,
as a function of the Coulomb interaction strength $U$.

In the weak-coupling regime, we found that 
the magnitude of the single-particle gap $\Delta_{\mr{gap}}$ is almost comparable to that of the order parameter $2|\Delta|$. 
This indicates that the electron-hole pair formation and its condensation may occur simultaneously,
like Cooper pair formation and its condensation in BCS theory.
The quasi-particle dispersion obtained by the single-particle excitation spectra $A(\mb{k},\w)$ is well described by mean-field theory.  
The Fermi momentum $\mb{k}_{\mr{F}}$ is defined from the momentum distribution function.
The anomalous excitation spectra $F(\mb{k},\w)$ showed that its spectral weight distributed mainly near the Fermi energy $\mu$.
The anomalous momentum distribution function $F(\mb{k})$ is peaked at $\mb{k}_\mr{F}$. 
Reflecting this, the coherence length of the exciton spread wide for several hundred lattice spacing.
This indicates that the system is in the BCS-like weakly-bound exciton condensation state.  
With increasing $U$, the momentum dependence of the condensation amplitude $F(\mb{k})$ becomes weak,
and the coherence length decreases rapidly.
$A(\mb{k},\w)$ show incoherent continua in their high-frequency part in the intermediate-coupling regime.  
In the strong-coupling regime, 
the energy scale of the order parameter and single-particle gap became separated.
This result indicates that the binding energy of the electron-hole pairs (excitons) is 
larger than the energy scale of the critical temperature
where exitons may obtain coherence.
The Fermi momentum $\mb{k}_{\mr{F}}$ became ill-defined. Accordingly, the Fermi surface plays no roles and  
$F(\mb{k})$ is widely spread in momentum space, and the coherence length is smaller than the lattice constant,
indicating that the system is in the BEC-like condensation state of strongly bound electron-hole pairs. 
Moreover, we found that HFA works well not only in the weak-coupling (small $U$) regime, but also in the strong-coupling (large $U$) regime.
The reason was clarified from the view point of the self-energy.
Finally, we discussed the spectral feature of the EI state of the EFKM and 
gave experimental implications for photoemission spectroscopy measurements.

\section*{Acknowledgment}
We thank H. Fehske, S. Ejima, T. Mizokawa, Y. Wakisaka, T. Konishi, T. Kaneko, and  S. Yamaki for useful discussions.  
KS acknowledges support from JSPS Research Fellowship for Young Scientists.  
This work was supported in part by Kakenhi Grant No.~22540363 of Japan.  
A part of computations was done at Research Center for Computational Science, Okazaki, Japan.

\section*{Appendix}
Here we derive the sum-rule for the self-energy used in Eq. (\ref{sumrule}). 
The EFKM or the asymmetric Hubbard model in momentum space is give as 
\begin{equation}
{\cal H} = 
\sum_{\mb{k} \s} \eps_{\mb{k} \s} c_{\mb{k} \s}^\dag c_{\mb{k} \s} + 
\frac{U}{L} \sum_{\mb{k} \mb{k}' \mb{q}} 
c_{\mb{k}  + \mb{q} \up}^\dag 
c_{\mb{k}           \up} 
c_{\mb{k}' - \mb{q} \dn}^\dag 
c_{\mb{k}'          \dn}.
\end{equation}

Here we consider the second-moment of the Green's function of the electron with spin $\s$ and momentum $\mb{k}$, which is  defined as~\cite{Harris} 
\begin{equation}\label{eq.M2G}
M^2_{\mb{k} \s} = \oint \frac{\dd \w}{2 \pi \imag} \w^2 G_\s(\mb{k},\w),  
\end{equation}
where $G_\s(\mb{k},\w)$ is the single-particle Green's function of the electron with spin $\s$, 
and the integral path encloses all singularities of the integrand. 
The Dyson equation gives the Green's function with the form 
\begin{equation}\label{eq.Gkw}
G_\s(\mb{k},\w) = \left(\w - \eps_{\mb{k}\s} - \Sigma_{\s}(\mb{k},\w) \right)^{-1}.
\end{equation}
Then we substitute the spectral representation of the self-energy \cite{Luttinger2}
\begin{equation}
\Sigma_\s(\mb{k},\w) = g_{\mb{k} \s} + \sum_\nu \frac{\sigma_{\mb{k} \s, \nu}} {\w - \zeta_{\mb{k} \s,\nu}}
\end{equation}
into (\ref{eq.Gkw}) and take the high-frequency expansion, 
\begin{eqnarray}
G_{\s} (\mb{k},\w) 
&=&
  \frac{1}{\w}
+ \frac{\eps_{\mb{k}\s} + g_{\mb{k}\s}}{\w^2} 
+ \frac{\sum_\nu \sigma_{\mb{k} \s,\nu} + \left(\eps_{\mb{k} \s}  + g_{\mb{k} \s} \right)^2}{\w^3} \notag  \\
&&
 + {\cal O} \left( \w^{-4} \right).
\end{eqnarray}
Substituting this expression into Eq.~(\ref{eq.M2G}) and using the theorem of residuum, we obtain
\begin{equation}\label{eq.M2-2}
M^2_{\mb{k} \s} =  \sum_\nu \sigma_{\mb{k} \s,\nu} + \left(\eps_{\mb{k} \s}  + g_{\mb{k} \s} \right)^2. 
\end{equation}

The other expression for the second moment is given as~\cite{Harris}
\begin{equation}
M^2_{\mb{k}\s}= \left \langle \left \{ [ c_{\mb{k}\s},{\cal H} ], [ {\cal H},c_{\mb{k}\s}^\dag ] \right \}_{+} \right \rangle,
\end{equation}
where $\{ \cdots \}_+$ denotes the anticommutator. Calculating the (anti) commutators on the right-hand-side, we obtain
\begin{equation} \label{eq.M2-1}
M^2_{\mb{k} \s} =  \eps_{\mb{k} \s}^2 + 2 \eps_{\mb{k} \s} U n_{\bar{\s}} + U^2 n_{\bar{\s}}, 
\end{equation}
where $\bar \s$ denotes the opposite spin direction of $\s$.
Now we use the fact that $g_{\mb{k} \s}$ is the Hartree potential,~\cite{Eder} i.e.,  $g_{\mb{k} \s} = U n_{\bar{\s}} $.
Then comparing Eq. (\ref{eq.M2-1}) with Eq. (\ref{eq.M2-2}), we obtain
\begin{equation}
\sum_\nu \sigma_{\mb{k} \s,\nu} = U^2   n_{\bar{\s}}  ( 1 -  n_{\bar{\s}})
\end{equation}
The right hand side does not depend on the momentum $\mb{k}$ or dispersion $\eps_{\mb{k} \s}$.
By replacing $\sigma = \up,\dn$ to $c,f$ and using the half-filling condition $n_c + n_f = 1$, 
we obtain Eq.~(\ref{sumrule}).


\begin{thebibliography}{99}
\bibitem{Mott} N. F. Mott,                               { Phil. Mag.}      {\bf  6},  287 (1961).
\bibitem{Halperin} B. I. Halperin, and T. M. Rice,       { Rev. Mod. Phys.} {\bf 40},  755 (1968).
\bibitem{Jerome} D. J\'{e}rome, T. M. Rice, and W. Kohn, { Phys. Rev.}      {\bf 158}, 462 (1967).
\bibitem{Cercellier} H. Cercellier, C. Monney, F. Clerc, C. Battaglia, L. Despont, M. G. Garnier, H. Beck, P. Aebi, L. Patthey,  H. Berger, and L. Forro,
                                                                          { Phys. Rev. Lett.}   {\bf 99}, 146403       (2007).
\bibitem{Monney} C. Monney, E. F. Schwier, M. G. Garnier, N. Mariotti, C. Didiot, H. Cercellier, J. Marcus, H. Berger, A. N. Titov, H. Beck, and P. Aebi, 
                                                                          { New J. Phys.}       {\bf 12}, 125019       (2010).  
\bibitem{Wakisaka} Y. Wakisaka, T. Sudayama, K. Takubo, T. Mizokawa, M. Arita, H. Namatame, M. Taniguchi, N. Katayama, M. Nohara, and H. Takagi,  
                                                                          { Phys. Rev. Lett.}   {\bf 103}, 026402      (2009).     
\bibitem{Kaneko} T. Kaneko, T. Toriyama, Y. Ohta, and T. Konishi, in preparation.                                                                          

\bibitem {FalicovKimball} L. M. Falicov and J. C. Kimball,                { Phys. Rev. Lett.}   {\bf 22}, 997          (1969).
                                                        
\bibitem {Batista1} C. D. Batista, J. E. Gubernatis, J. Bonca, and H. Q. Lin, 
                                                                          { Phys. Rev. Lett.}   {\bf 92}, 187601       (2004).
\bibitem {Batista2} C. D. Batista,                                        { Phys. Rev. Lett.}   {\bf 89}, 166403       (2002).

\bibitem {Zenker1} B. Zenker, D. Ihle, F. X. Bronold, and H. Fehske,       { Phys. Rev. B}       {\bf 83}, 235123       (2011).
\bibitem {Farkasovsky} P. Farkasovsky,                 { Phys. Rev. B}       {\bf 77}, 155130       (2008). 
\bibitem {Ihle} D. Ihle, M. Pfafferott, E. Burovski, F. X. Bronold, and H. Fehske,  { Phys. Rev. B}       {\bf 78}, 193103       (2008).
                

\bibitem {Phan1} V. N. Phan, K. W. Becker, and H. Fehske,                 { Phys. Rev. B}       {\bf 81}, 205117       (2010).
\bibitem {Phan2} V. N, Phan, H. Fehske, and K. W. Becker,                 { EPL}                {\bf 95}, 17006        (2011).  
\bibitem {Bronold} F. X. Bronold and H. Fehske,                           { Phys. Rev. B}       {\bf 74}, 165107       (2006).
\bibitem{Potthoff1}  M. Potthoff, M. Aichhorn, and C. Dahnken,            { Phys. Rev. Lett.}   {\bf 91}, 206402       (2003).
\bibitem{Potthoff2}  M. Potthoff,                                         { Eur. Phys. J. B}    {\bf 32}, 429 (2003); {\bf 36}, 335 (2003).
\bibitem{Senechal2} D. S\'{e}n\'{e}chal, D. Perez, and M. Pioro-Ladriere, { Phys. Rev. Lett.}   {\bf 84},  522       (2000).  

\bibitem{Senechal3} A. H. Nevidomskyy, C. Scheiber, D. S\'{e}n\'{e}chal, and A.-M. S. Tremblay,  { Phys. Rev. B} {\bf 77}, 064427 (2008)     
\bibitem{Yoshikawa} T. Yoshikawa and M. Ogata,                          { Phys. Rev. B} {\bf 79}, 144429 (2009)
\bibitem{Horiuchi} S. Horiuchi, S. Kudo, T. Shirakawa, and Y. Ohta,      { Phys. Rev. B} {\bf 78}, 155128 (2008) 
\bibitem{Watanabe} H. Watanabe, T. Shirakawa, and S. Yunoki,            { Phys. Rev. Lett.} {\bf 105}, 216410 (2010)
                                                                                              
\bibitem{Nozieres} P. Nozi\`{e}res and S. Schmitt-Rink,                   { J. Low Temp. Phys.} {\bf 59}, 195          (1985).
\bibitem{Luttinger}  J. M. Luttinger and J. C. Ward,                      { Phys. Rev.}         {\bf 118}, 1417        (1960).
\bibitem{Aichhorn}  M. Aichhorn, E. Arrigoni, M. Potthoff, and W. Hanke,  { Phys. Rev. B}       {\bf 74},  024508      (2006).
\bibitem{Potthoffreview} M. Potthoff,                                     { arXiv}                         1108:2183v1 (2011).
\bibitem{Senechalreview} D. S\'{e}n\'{e}chal,                             { arXiv}                         0806:2690v2 (2008).

                                                            
\bibitem{Toschi} A. Toschi, M. Capone, and C. Castellani,                 { Phys. Rev. B}       {\bf 72}, 235118       (2005).    
\bibitem{Koga} A. Koga and P. Werner,                                     { Phys. Rev. A}       {\bf 84}, 023638       (2011).
\bibitem{Ohta} Y. Ohta, A. Nakauchi, R. Eder, K. Tsutsui, and S. Maekawa, { Phys. Rev. B}       {\bf 52}, 15617        (1995).
\bibitem{Luttinger2}     J. M. Luttinger,                                 { Phys. Rev.}         {\bf 121}, 942         (1961).
\bibitem{Turkowski} V. Turkowski and J. K. Freericks,                   { Phys. Rev. B} {\bf 77 }, 205102 (2008). 

\bibitem{MerminWagner} N. D. Mermin and H. Wagner,                      {Phys. Rev. Lett} {\bf 17}, 1133 (1966).
\bibitem{Hohenberg} P. C. Hohenberg,                                    {Phys. Rev.}      {\bf 158}, 383 (1967).
\bibitem {Zenker2} B. Zenker, D. Ihle, F. X. Bronold, and H. Fehske,       { Phys. Rev. B}       {\bf 81}, 115122       (2010).
\bibitem{Damascelli} A. Damascelli, Z. Hussain, and Z.-X. Shen,          { Rev. Mod. Phys.} {\bf 75}, 473 (2003).

\bibitem{Katagiri} D. Katagiri, K. Seki, R. Eder, and Y. Ohta,           { Phys. Rev. B} {\bf 83}, 165124 (2011)
\bibitem{Yeh} J. J. Yeh and I. Lindau,                                   { At. Data Nucl. Data Tables} {\bf 32}, 1 (1985).
\bibitem{Eder} R. Eder, K. Seki, and Y. Ohta,                           { Phys. Rev. B} {\bf 83}, 205137 (2011)
\bibitem{Harris} A. B. Harris and R. V. Lange,                          { Phys. Rev.} {\bf 157}, 295 (1967).
\end{thebibliography}
\end{document}